\documentclass[conference,a4paper]{IEEEtran}
\usepackage{cite}
\usepackage{amsmath,amssymb,amsfonts}
\usepackage{algorithmic}
\usepackage{graphicx}
\usepackage{textcomp}
\usepackage{multirow}
\usepackage{array}
\usepackage{makecell}
\usepackage{xcolor}
\usepackage{subcaption}
\usepackage{booktabs}
\usepackage{tabularx}
\usepackage{tabularx}
\usepackage{booktabs}
\usepackage{multirow}
\usepackage{colortbl}
\usepackage{xcolor}
\definecolor{lightgray}{gray}{0.9} % Light gray for header shading
\newcolumntype{Y}{>{\centering\arraybackslash}X}

\ifCLASSOPTIONcompsoc
 \usepackage[caption=false,font=normalsize,labelfont=sf,textfont=sf]{subfig}
\else
 \usepackage[caption=false,font=footnotesize]{subfig}
\fi
\begin{document}
\title{Measurement-Based Analysis of Outdoor Massive MIMO Channel Characteristics over FR3 Frequency Band}

\author{
\IEEEauthorblockN{
Enrui Liu\IEEEauthorrefmark{1},
Pan Tang\IEEEauthorrefmark{1}, 
Haiyang Miao\IEEEauthorrefmark{1},
Qi Zhen\IEEEauthorrefmark{1},
Jianhua Zhang\IEEEauthorrefmark{1},
and Sen Wang\IEEEauthorrefmark{2}
}
\IEEEauthorblockA{\IEEEauthorrefmark{1}Beijing University of Posts and Telecommunications, Beijing, China\\
\IEEEauthorblockA{\IEEEauthorrefmark{2}China Mobile Research Institute, Beijing, China\\
Email: \{liuenrui, tangpan27, hymiao, zq2024018002, jhzhang\}@bupt.edu.cn}
wangsenyjy@chinamobile.com}
}



\maketitle

\begin{abstract}
The Frequency Range 3 (FR3) band is attracting increasing attention due to limited lower-frequency spectrum and growing mobile communication demand. This study experimentally investigates channel characteristics in Urban Macro (UMa) scenarios at 8 GHz and 15 GHz using a large-scale MIMO platform with time-division multiplexing (TDM). Key parameters, including root mean square (RMS) delay spread (DS) and angular spread (AS), were extracted and compared with 3rd Generation Partnership Project (3GPP) TR 38.901. Results reveal clear frequency-dependent behaviors: RMS delay spread remains nearly constant under line of sight (LOS) but decreases from 8 GHz to 15 GHz in non-line of sight (NLOS), indicating reduced multipath dispersion at higher frequencies. Both azimuthal spreads (including ASA and ASD) and elevation spreads (including ESA and ESD) exhibit a corresponding decrease with increasing frequency, demonstrating a consistent trend towards more directional propagation across all angular domains. Capacity analysis indicates that the 15 GHz channel slightly outperforms 8 GHz in both LOS and NLOS scenarios due to more concentrated multipath energy and larger dominant singular values. Higher frequencies exhibit greater directionality, whereas lower frequencies provide broader multipath distributions and more stable performance, offering valuable guidance for multi-band MIMO modeling and 6G system design.
\end{abstract}

\vskip0.5\baselineskip

% \fi
% \IEEEpeerreviewmaketitle

\section{Introduction}
The FR3 band, covering mid-range frequencies from 3.3 to 15 GHz, has received growing attention due to increasing demand for mobile data and limited spectrum at lower frequencies\cite{wrc23}\cite{FR3}. These bands provide substantial bandwidth with moderate propagation loss, making them suitable for urban macro cell and indoor coverage scenarios\cite{1}\cite{2}. Meanwhile, Massive MIMO technology\cite{6G}, through large-scale antenna arrays, can exploit these mid-band frequencies for enhanced spatial multiplexing and beamforming, improving spectral efficiency and system robustness. Investigating channel characteristics in the FR3 band is therefore crucial for accurate modeling, efficient network planning, and the design of future 6G systems.

To date, extensive research has been conducted on massive MIMO at multiple frequency bands. In \cite{haiyang}, channel measurements and analyses were carried out for the 6 GHz band in the UMa scenario, with further investigation into channel capacity characteristics. In \cite{haiyang2}, measurements were performed in UMi and O2I scenarios at 3.3, 6.5, 15, and 28 GHz, where the channel characteristics across multiple frequency bands were analyzed, and their impacts on wireless system performance were evaluated \cite{haiyang3}, including cell coverage radius, data rate, and bit error rate (BER). In channel measurements were presented at 4.9 and 28 GHz in indoor industrial scenarios, showing that the RMS delay spread (DS) at 28 GHz is about 50 ns smaller than that at 4.9 GHz. In \cite{peter}, the collected data at 60 GHz and 10 GHz were uniformly analyzed, with a particular focus on the frequency dependence of path loss (PL) and delay spread. Overall, previous studies provide valuable insights into MIMO channel characteristics but lack systematic investigations and large-scale measurements in the FR3 frequency range. However, many existing FR3 studies focus on frequencies close to the lower (around 6~GHz) or upper (15~GHz) boundaries of the band, while comparative large-scale measurement data around 8~GHz remain relatively limited. From a system perspective, 8~GHz represents a characteristic operating point within the upper mid-band, offering a distinct trade-off between array compactness, propagation richness, and coverage capability when compared with higher FR3 frequencies.

This study provides a comparative analysis of Massive MIMO propagation at 8 GHz and 15 GHz to characterize frequency-dependent trends within the FR3 band. To ensure high-fidelity data, hardware impairments (e.g., power fluctuations and antenna coupling) were mitigated through rigorous calibration, while measurement repeatability was guaranteed via multi-snapshot acquisition and synchronized antenna switching. These methodologies establish a robust baseline for 6G system-level evaluations. The main contributions are:
\begin{enumerate} \setlength{\itemsep}{0ex}

\setlength{\parskip}{0.5ex} \item We conduct large-scale Massive MIMO measurements at 8 GHz and 15 GHz in UMa scenarios and provide a more granular understanding of frequency-dependent trends in the FR3 range. \item We evaluate key multipath parameters, including RMS delay spread and angular spread, for both LOS and NLOS conditions. To ensure repeatability, measurements were performed across multiple time-stamps and locations, with a rigorous analysis of the consistency of the results. \item Using the measured channel matrices, we calculate and compare the spatial multiplexing capabilities of both bands, providing insights into the spectral efficiency limits of the 8 GHz and 15 GHz massive MIMO systems. \end{enumerate}

\section{Channel Measurement Platform and Environment}
\subsection{Measurement Platform}
\begin{figure}[!ht]
\begin{subfigure}{\columnwidth}
    \centering
    \includegraphics[scale=0.5]{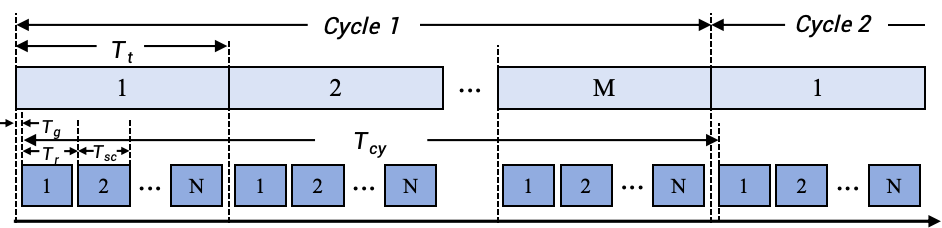}
    \caption{The principle of time-sequential switching}
    \label{fig:principle}
\end{subfigure}
\begin{subfigure}{\columnwidth}
    \centering
    \includegraphics[scale=0.25]{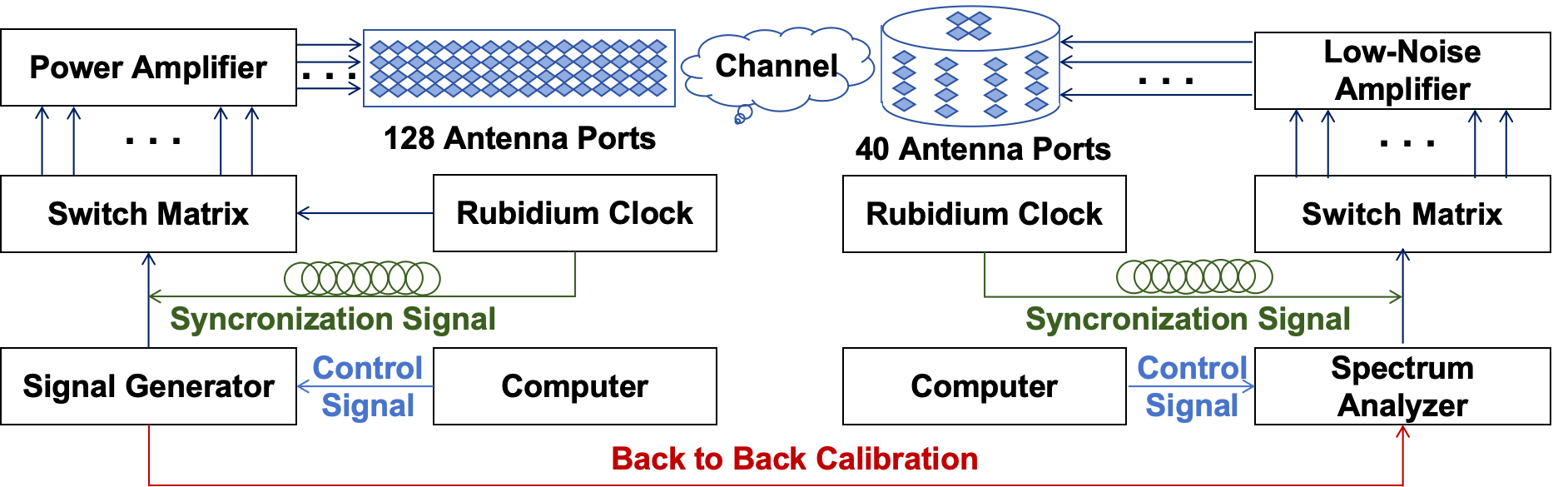}
    \caption{The top view of the measurement platform}
    \label{fig:platform}
\end{subfigure}
\caption{Schematic diagram of measurement platform.}
\label{Fig.1}
\end{figure}

The measurement platform utilizes a Time-Division Multiplexing (TDM) architecture, as illustrated in Fig. \ref{Fig.1}, with parameters detailed in Table \ref{Table.1}. The system employs a PN9 sequence with a 98 kHz repetition rate (10.22 $\mu s$ duration) for channel sounding. To capture the 128×40 MIMO channel, both Tx and Rx antenna arrays are interfaced through high-speed RF switches controlled by a host computer. For each active Tx element, 40 Rx ports are sequentially polled within the channel coherence time. By maintaining stationary transceivers, the wireless channel remains quasi-static during each snapshot, preventing temporal decorrelation and ensuring the integrity of the measured data. The signal flow is as follows: At the transmitter, a BPSK-modulated baseband signal is amplified and routed through a 128-channel switch matrix. At the receiver, the captured signals are enhanced by low-noise amplifiers (LNAs) and recorded as complex I/Q samples via a spectrum analyzer. Channel Impulse Responses (CIRs) are subsequently extracted through offline matched filtering. To ensure phase stability and nanosecond-level precision, both ends are synchronized using GPS-disciplined rubidium clocks.
\begin{table}[ht]
\begin{center}
\caption{Measurement system parameters.}
\label{Table.1}
\begin{tabular}{c|c}
\hline

\textbf{Parameters} & \textbf{Value}\\
\hline
Center Frequency & $3$-$16$ GHz\\
Maximum Band Width & $2$ GHz\\
number of Tx switch Matrix channel & $128$\\
number of Rx switch Matrix channel & $64$\\
Tx switch gain & $\geq 27$ dBi\\
Rx switch gain & $\geq 33$ dBi\\
Antenna angular resolution & $\leq 1 ^\circ$\\
Synchronous mode & GPS rubidium clock\\
\hline

\end{tabular}
\end{center}
\end{table}

In Fig. \ref{Fig.1} (a), $T_{t}$ denotes the transmission switching duration, i.e., the time during which the signal is maintained on the Tx antenna array; $T_{r}$ denotes the reception switching duration; $T_{cy}$ is the probing cycle time, which must satisfy the condition $T_{cy} \geq M T_t$; $T_g$ is the guard interval, which is reset to zero at the beginning of the next cycle; $T_{sc}$ represents the continuous detection duration of each Rx antenna element; and $T_s$ is the pulse period, during which each Rx antenna array performs one detection. The specified probing duration $T_t$ exactly corresponds to one detection cycle, i.e., $T_t$ = N$T_{sc}$. Since switching requires additional time in practical operation, a guard interval must be introduced. Therefore, the interval between two consecutive detection signals is defined as $T_r$, with the condition $T_r \geq T_{sc}$.

The entire measurement platform employs a PN-based channel sounding scheme. At the transmitter side, PN9 sequence codewords are generated using a signal generator and mapped onto a BPSK-modulated baseband waveform, thereby preserving both amplitude and phase information. The modulated signal is transmitted via coaxial cables to a matrix switching unit. The switching sequence is controlled by a host computer, allowing the 128 transmit channels to be sequentially connected to the RF chain, pass through power amplifiers, and excite the corresponding antenna ports for over-the-air transmission. At the receiver side, the signals captured by the antenna array are first amplified by low-noise amplifiers (LNAs) and then routed through RF switches to a spectrum analyzer operating in I/Q acquisition mode. The received complex baseband I/Q samples are recorded and subsequently processed offline to extract the channel impulse responses (CIRs) through matched filtering with the known PN sequence. The timing and synchronization of the entire measurement platform are governed by a GPS-disciplined rubidium clock system, which provides a common frequency and time reference for both the transmitter and receiver. This ensures nanosecond-level synchronization accuracy across all transmit and receive channels, thereby enabling reliable and high-precision massive MIMO channel measurements.
\subsection{Measurement Environment}
\begin{figure}[!ht]
\centering
\begin{subfigure}{0.45\columnwidth}
    \centering
    \includegraphics[width=\textwidth,height=0.125\textheight]{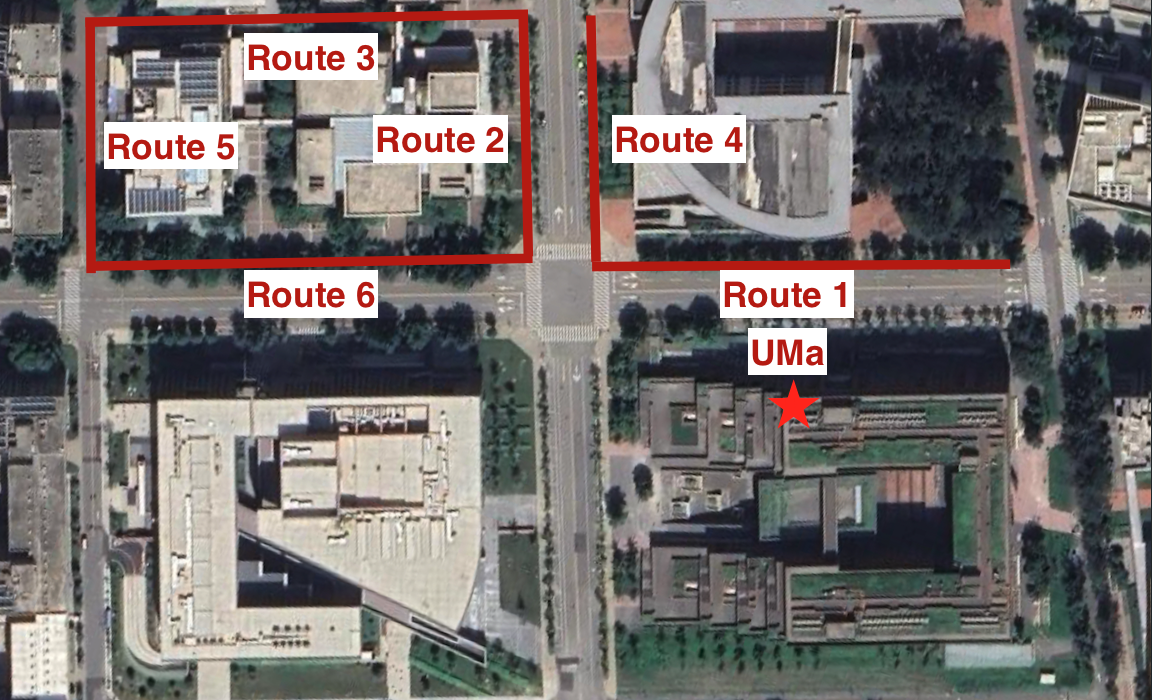}
    \caption{Route schematic}
    \label{fig:route}
\end{subfigure}
\hspace{0.05\columnwidth}
\begin{subfigure}{0.45\columnwidth}
    \centering
    \includegraphics[width=\textwidth]{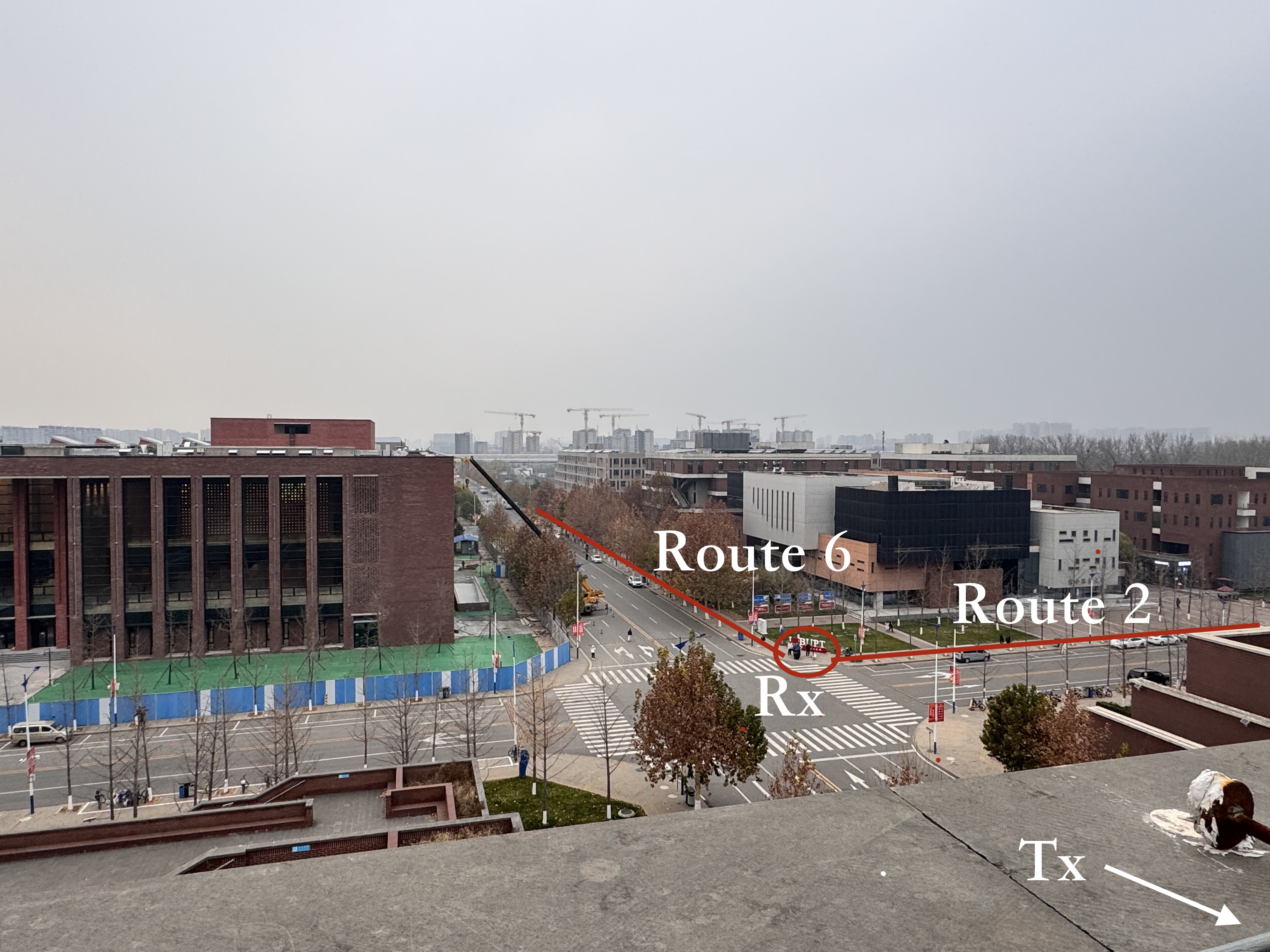}
    \caption{View from transmitter}
    \label{fig:tx_view}
\end{subfigure}
\\[0.5em]
\begin{subfigure}{0.45\columnwidth}
    \centering
    \includegraphics[width=\textwidth]{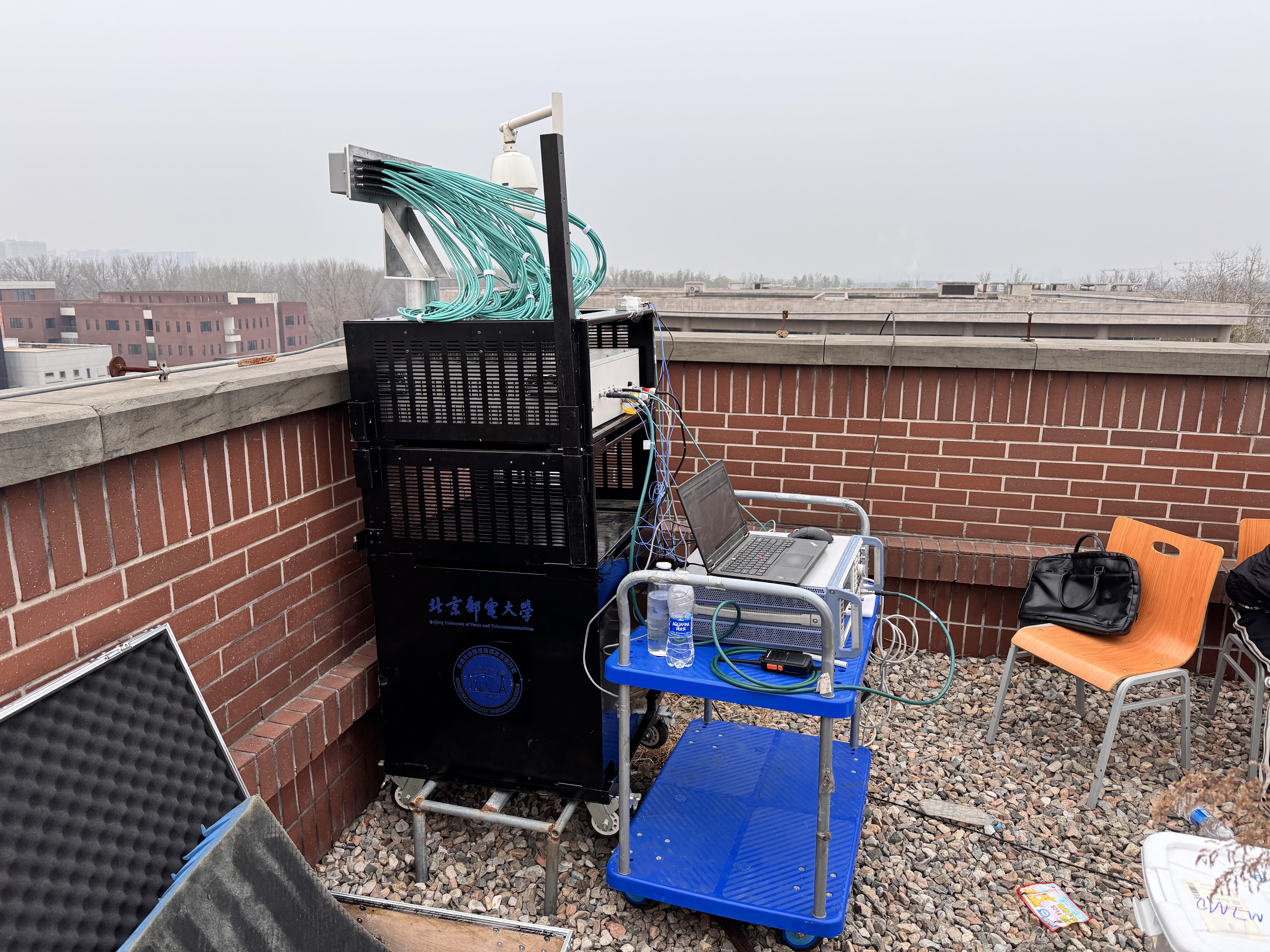}
    \caption{Photo of transmitter}
    \label{fig:transmitter}
\end{subfigure}
\hspace{0.05\columnwidth}
\begin{subfigure}{0.45\columnwidth}
    \centering
    \includegraphics[width=\textwidth]{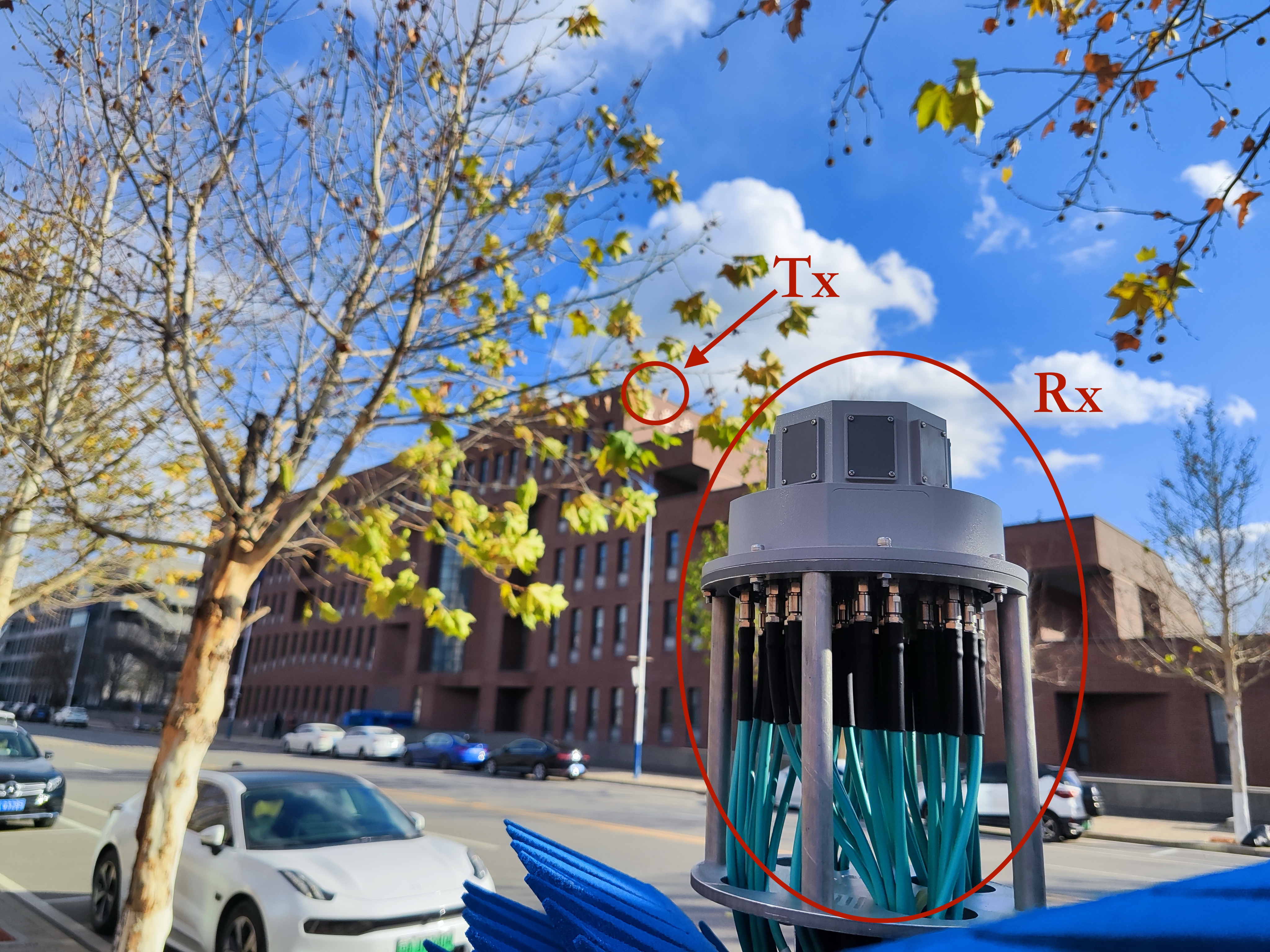}
    \caption{Photo of receiver}
    \label{fig:receiver}
\end{subfigure}
\caption{Measurement Setup and Hardware Overview.}
\label{Fig.2}
\end{figure}

The measurement campaign was carried out on the campus of Beijing University of Posts and Telecommunications (BUPT), Shahe Campus. For the UMa scenario, the transmitter was installed on the rooftop of the BUPT main building, approximately 27.8 m above ground, as marked by the stars in Fig. \ref{Fig.2} (a). The receiver was moved along the routes illustrated in Fig. \ref{Fig.2}(a), covering six routes with approximately 12 uniformly distributed measurement points on each route. At each measurement point, the receiver was kept stationary during the data acquisition process to ensure quasi-static channel conditions. The photos of transmitter and receiver are shown in Fig. \ref{Fig.2} (c) and \ref{Fig.2} (d), and the transmitter’s perspective is depicted in Fig. \ref{Fig.2} (b).

\subsection{Measurement Data Post-Processing}
A PN9 sequence is employed for its superior autocorrelation properties and wide dynamic range. To isolate the true channel impulse response (CIR) from the inherent system response, a direct cable calibration is performed prior to the measurement campaign \cite{wang}\cite{tangpan}. The received signal $y(t)$ is expressed as:
\begin{equation}
y(\tau) = x(\tau)*h_{Tx}(\tau)*a_{Tx}(\tau)*h(\tau)*a_{Rx}(\tau)*h_{Rx}(\tau),
\end{equation}
where “*” denotes the convolution operation. Here, $x(\tau)$ is the transmitted PN sequence, $h_{Tx}(\tau)$ and $h_{Rx}(\tau)$ represent the responses of the transmitter and receiver RF chains, $a_{Tx}(\tau)$ and $a_{Rx}(\tau)$ denote the characteristics of the transmit and receive antennas, and $h(\tau)$ is the channel impulse response to be measured. The calibration signal is expressed as
\begin{equation}
y_{cal}(\tau) = x(\tau)*h_{Tx}(\tau)*h_{Rx}(\tau),
\end{equation}
by performing calibration, the effects of the transceiver RF chains can be removed. Specifically, the effective system response $h_{sys}(\tau) = h_{Tx}(\tau)*h_{Rx}(\tau)$ is first obtained through the direct connection measurement. Then, the received signal $y(\tau)$ is divided by the calibration response $y_{cal}(\tau)$ in the frequency domain, yielding
\begin{equation}
H(\tau) = \mathcal{F}^{-1}\left \{  {\frac{Y(f)}{Y_{cal}(f)}}\right \}  = a_{Tx}(\tau)*h(\tau)*a_{Rx}(\tau),
\end{equation}
where $\mathcal{F}^{-1}{\cdot}$ denotes the inverse Fourier transform, and $Y(f)$ and $Y_{cal}(f)$ are the Fourier transforms of $y(\tau)$ and $y_{cal}(\tau)$, respectively.

In this way, the influence of the RF front-end is eliminated, and the obtained result contains only the channel impulse response combined with the antenna characteristics. If necessary, further de-embedding of antenna effects can be applied to isolate the pure channel response $h(\tau)$. After calibration and extraction of the channel impulse response, the high-resolution Space-Alternating Generalized Expectation-Maximization (SAGE) algorithm is employed for parameter estimation\cite{sage}. 

\section{Channel Characterisation and Modeling}
In both LOS and NLOS scenarios, the presence of multiple scatterers leads to different propagation behaviors and frequency-dependent characteristics of electromagnetic waves. Therefore, this section models and analyzes key channel characteristics and comparatively investigates the influence of frequency and propagation conditions on signal quality.

\subsection{RMS Delay Spread}
The RMS-DS is a fundamental parameter for characterizing the temporal dispersion of a wireless channel. It quantifies the spread of multipath components in the time domain and describes how signal energy is distributed around the mean arrival time.

Given the power delay profile (PDP) $P(\tau_i)$ with delays $\tau_i$, the mean delay $\tau_\text{mean}$ and RMS Delay Spread $\tau_\text{rms}$ are defined as:
\begin{equation}
\tau_{mean} = \frac{\sum_{l=1}^L P(\tau_l)\tau_l}{\sum_{l=1}^L P(\tau_l)},
\end{equation}
\begin{equation}
\tau_{rms}=\sqrt{\frac{\sum_{l=1}^{L}{(\tau_{l}-\tau_{mean})^2}P(\tau_l)}{\sum_{l=1}^LP(\tau_l)}},
\end{equation}
where, $\tau_\text{mean}$ represents the average arrival time of the multipath components, while $\tau_\text{rms}$ indicates how these components are dispersed around the mean. A larger RMS-DS corresponds to stronger temporal dispersion, which may lead to inter-symbol interference (ISI). Conversely, a smaller RMS-DS suggests that multipath components are more temporally concentrated, resulting in a more stable received signal.

\begin{figure}[!ht]
\centering
\begin{subfigure}{0.45\columnwidth}
    \centering
    \includegraphics[width=\textwidth]{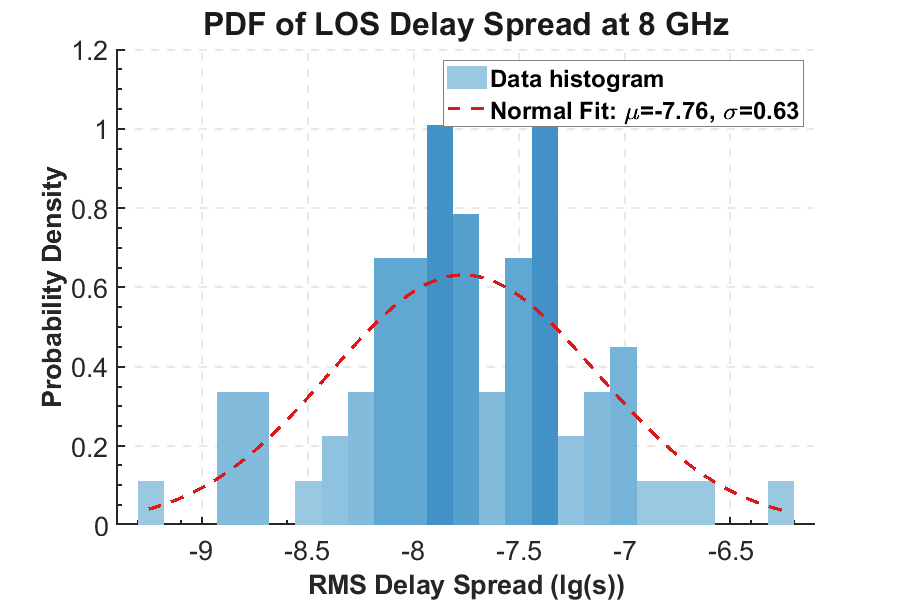}
    \caption{}
    \label{fig:los_8ghz}
\end{subfigure}
\hspace{0.05\columnwidth}
\begin{subfigure}{0.45\columnwidth}
    \centering
    \includegraphics[width=\textwidth]{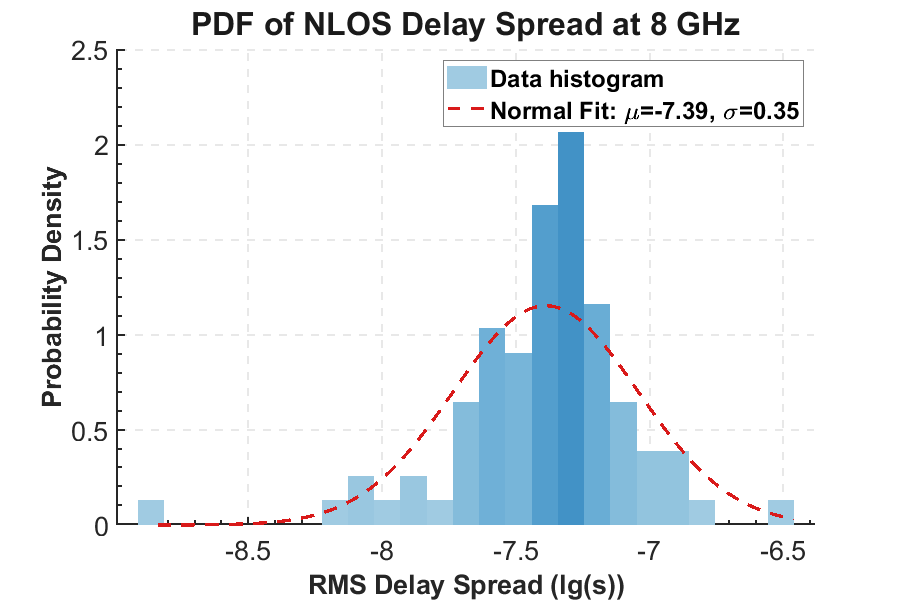}
    \caption{}
    \label{fig:nlos_8ghz}
\end{subfigure}
\\[0.5em]
\begin{subfigure}{0.45\columnwidth}
    \centering
    \includegraphics[width=\textwidth]{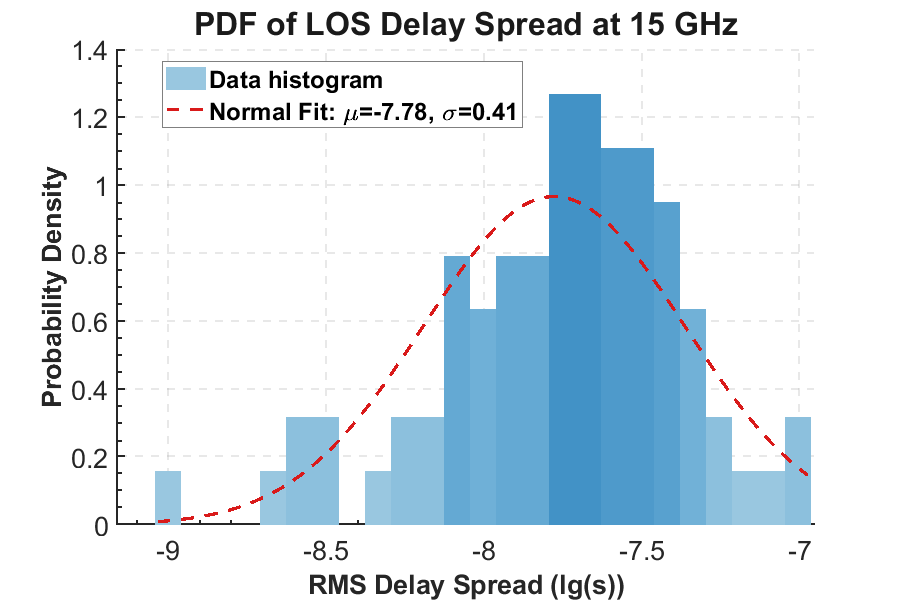}
    \caption{}
    \label{fig:los_15ghz}
\end{subfigure}
\hspace{0.05\columnwidth}
\begin{subfigure}{0.45\columnwidth}
    \centering
    \includegraphics[width=\textwidth]{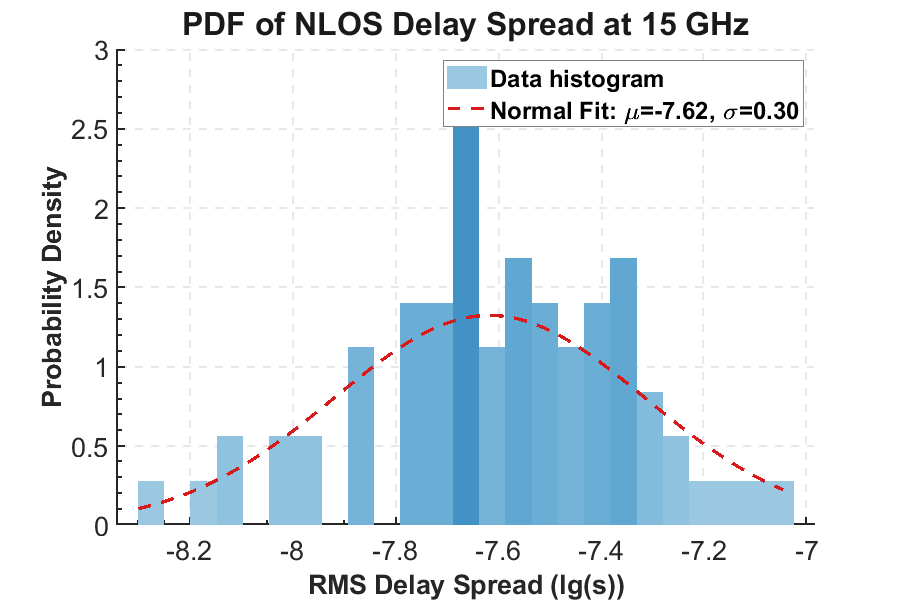}
    \caption{}
    \label{fig:nlos_15ghz}
\end{subfigure}
\caption{RMS delay spread results under different frequency bands and propagation scenarios. 
(a) 8\,GHz LOS scenario; 
(b) 8\,GHz NLOS scenario; 
(c) 15\,GHz LOS scenario; 
(d) 15\,GHz NLOS scenario.}
\label{Fig.Delay}
\end{figure}

Fig.~\ref{Fig.Delay} and TABLE.~\ref{tab:delay} illustrates the measured RMS Delay Spread under LOS and NLOS conditions at different frequencies.
The measured RMS delay spread demonstrates markedly different sensitivities to carrier frequency in LOS and NLOS scenarios. For LOS propagation, the results at 8 GHz (–7.76, 17.38 ns) and 15 GHz (–7.78, 16.61 ns) are remarkably consistent, showing minimal frequency dependence. However, a clear frequency dependence emerges in the NLOS case, where the delay spread decreases substantially from 40.7 ns at 8 GHz to 23.9 ns at 15 GHz.This reduction in temporal dispersion at 15 GHz under NLOS conditions indicates that higher-frequency signals experience a less rich multipath environment. The higher path loss and increased scattering loss at 15 GHz likely attenuate distant multipath components, leading to a tighter channel impulse response.

In conclusion, while the LOS/NLOS discrepancy is persistent, the temporal characteristics in NLOS environments are significantly more sensitive to the operating frequency. This finding underscores that system design, particularly for NLOS applications, must carefully account for the impact of frequency choice on delay spread.

\begin{table}[t]
    \centering
    \caption{Statistical Parameters of RMS Delay Spread}
    \renewcommand{\arraystretch}{1.25} % 增加行距
    \setlength{\tabcolsep}{10pt} % 列间距
    \begin{tabularx}{0.95\columnwidth}{l *{4}{>{\centering\arraybackslash}X}}
        \toprule
        \multirow{2}{*}{\textbf{Parameter}} & \multicolumn{2}{c}{\textbf{8 GHz}} & \multicolumn{2}{c}{\textbf{15 GHz}} \\
        \cmidrule(lr){2-3} \cmidrule(lr){4-5}
        & \textbf{LOS} & \textbf{NLOS} & \textbf{LOS} & \textbf{NLOS} \\
        \midrule
        $\mu$     & $-7.76$ & $-7.39$ & $-7.78$ & $-7.62$ \\
        $\sigma$  & $0.63$ & $0.35$ & $0.41$ & $0.30$ \\
        \bottomrule
    \end{tabularx}
    \label{tab:delay}
\end{table}
\subsection{RMS Angle Spread}
While RMS Delay Spread captures the temporal dispersion of multipath components, the RMS-AS characterizes their spatial dispersion. Quantifies how the power of the multipath components is distributed around the mean arrival or departure angle, thus describing the angular richness of the channel.

For multipath components with angles $\theta_i$ and powers $P(\theta_i)$, the mean angle $\theta_\text{mean}$ and RMS Angle Spread $\theta_\text{rms}$ are expressed as \cite{zhang}:
\begin{equation}
\theta_\text{mean} = \frac{\sum_{i=1}^{N} P(\theta_i) \theta_i}{\sum_{i=1}^{N} P(\theta_i)},
\end{equation}
\begin{equation}
\theta_\text{rms} = \sqrt{\frac{\sum_{i=1}^{N} P(\theta_i) (\theta_i - \theta_\text{mean})^2}{\sum_{i=1}^{N} P(\theta_i)}},
\end{equation}
where, $\theta_\text{mean}$ denotes the average angle of arrival or departure, and $\theta_\text{rms}$ represents the angular spread around this mean. A larger RMS-AS implies that multipath components arrive from a wider range of directions. In contrast, a smaller RMS-AS indicates that the energy received is concentrated within a narrow angular region, facilitating directional transmission and improving array efficiency.

\begin{table*}[t]
\small % Slightly smaller font for better fit
\centering
\caption{Comparison of Angular Spread Parameters between Measurement and 3GPP}
\label{table:angular_spread}
\begin{tabularx}{\textwidth}{c|c|Y|Y|Y|Y|Y|Y|Y|Y}
\toprule
\rowcolor{lightgray}
\multicolumn{2}{c|}{} 
  & \multicolumn{4}{c|}{\textbf{Measurement}} 
  & \multicolumn{4}{c}{\textbf{3GPP}} \\
\cmidrule(lr){3-6}\cmidrule(lr){7-10}
\rowcolor{lightgray}
\multicolumn{2}{c|}{\textbf{Angular Spread ($\log_{10}[^\circ]$)}}
  & \multicolumn{2}{c|}{\textbf{8 GHz}} & \multicolumn{2}{c|}{\textbf{15 GHz}}
  & \multicolumn{2}{c|}{\textbf{8 GHz}} & \multicolumn{2}{c}{\textbf{15 GHz}} \\
\cmidrule(lr){3-4}\cmidrule(lr){5-6}\cmidrule(lr){7-8}\cmidrule(lr){9-10}
\rowcolor{lightgray}
\multicolumn{2}{c|}{}    & \textbf{LOS} & \textbf{NLOS} & \textbf{LOS} & \textbf{NLOS} & \textbf{LOS} & \textbf{NLOS} & \textbf{LOS} & \textbf{NLOS} \\
\midrule
\multirow{2}{*}{\textbf{ASA}} & $\mu$     & 1.80 & 1.86 & 1.57 & 1.70 & 1.81 & 1.84 & 1.81 & 1.76 \\
                              & $\sigma$  & 0.21 & 0.12 & 0.24 & 0.24 & 0.20 & 0.11 & 0.20 & 0.11 \\
\midrule
\multirow{2}{*}{\textbf{ASD}} & $\mu$     & 1.25 & 1.32 & 1.24 & 1.21 & 1.16 & 1.40 & 1.19 & 1.37 \\
                              & $\sigma$  & 0.21 & 0.15 & 0.27 & 0.23 & 0.28 & 0.28 & 0.28 & 0.28 \\
\midrule
\multirow{2}{*}{\textbf{ESA}} & $\mu$     & 1.20 & 1.19 & 1.13 & 1.17 & 0.95 & 1.22 & 0.95 & 1.31 \\
                              & $\sigma$  & 0.29 & 0.24 & 0.29 & 0.30 & 0.16 & 0.16 & 0.16 & 0.16 \\
\midrule
\multirow{2}{*}{\textbf{ESD}} & $\mu$     & 1.33 & 1.23 & 1.22 & 1.14 & -- & -- & -- & -- \\
                              & $\sigma$  & 0.17 & 0.25 & 0.25 & 0.19 & -- & -- & -- & -- \\
\bottomrule
\end{tabularx}
\end{table*}
As summarized in Table~\ref{table:angular_spread}, the measured AS parameters at 8 GHz and 15 GHz systematically deviate from the 3GPP TR 38.901 reference values, particularly in the elevation domain. These deviations are primarily caused by frequency-dependent interactions with the UMa environment. At 15 GHz, the shorter wavelength promotes diffuse scattering; however, the higher material absorption simultaneously suppresses weaker high-order reflections. This leads to a "spatial thinning" effect, where received power is concentrated in fewer, more dominant clusters compared to the 8 GHz band. Furthermore, the significantly higher measured ESA (e.g., 1.20 at 8 GHz vs. 0.95 in 3GPP) indicates that the standard model underestimates vertical scattering from high-rise structures in dense UMa scenarios.

To better align the 3GPP framework with these FR3 characteristics, the frequency-scaling laws for the 7--15 GHz range should be recalibrated. Specifically, the $\mu lgAS$ offset requires adjustment to reflect the steeper reduction in azimuth spread at higher frequencies, and the mean elevation spread parameters should be increased to capture the richer vertical multipath observed in real-world measurements. These refinements ensure that future 6G spatial multiplexing and beamforming strategies are evaluated against a more physically representative channel baseline.

\subsection{Channel Capacity}

The channel capacity \( C \) \cite{capacity} in a MIMO system can be expressed as:

\begin{equation}
C = \log_2 \det \left( I_{N_R} + \frac{\text{SNR}}{N_T} H H^H \right),
\end{equation}
where \( I_{N_R} \) is the identity matrix with dimensions corresponding to the number of receive antennas \( N_R \), \( N_T \) is the number of transmit antennas, \( H \) is the channel matrix, \( H^H \) is its conjugate transpose and SNR is the signal-to-noise ratio. The characteristics of the channel matrix \( H \) are influenced by the angle spread, with the array response vector \( a(\theta, \phi) = \exp \left( j \pi (0:N-1) \sin(\theta) \cos(\phi) \right) \) describing the angular distribution.  
\begin{figure}[!ht]
\centering
    \includegraphics[width=0.8\columnwidth]{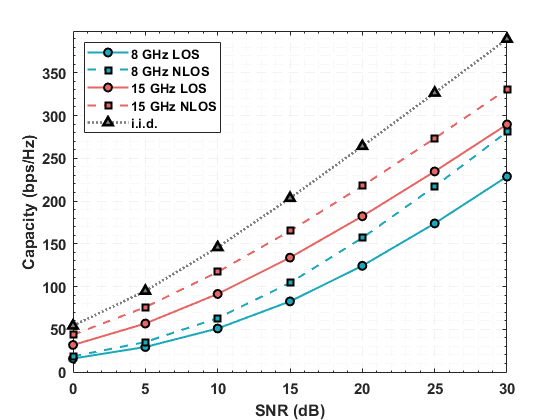}
\caption{Comparative Analysis of Channel capacity across 8 GHz and 15 GHz Frequency Bands.}
\label{Fig.3}
\end{figure}

This study compares MIMO channel capacities at 8 GHz and 15 GHz under LOS and NLOS conditions to evaluate the impact of frequency on propagation performance, as shown in Fig. \ref{Fig.3}. The results indicate that the 15 GHz band achieves higher capacity than 8 GHz across the evaluated SNR range. Although the 15 GHz channel exhibits smaller delay and angular spreads—reflecting a sparser multipath environment—its energy is more concentrated in dominant propagation paths. This results in a channel matrix with larger initial singular values, signifying higher effective channel gain in the primary spatial layers. In contrast, while the 8 GHz channel features richer scattering, its energy is distributed more broadly across weaker paths, yielding smaller dominant singular values. In the considered scenarios, the gain from concentrated power at 15 GHz outweighs the spatial multiplexing benefits provided by the richer multipath at 8 GHz, leading to superior channel efficiency.
\section{Conclusion}
The measurement results reveal clear frequency-dependent behaviors in the delay, angular, and spatial domains. In LOS conditions, the RMS delay spread remains nearly constant across frequencies, while in NLOS it decreases from 8 GHz to 15 GHz, indicating reduced multipath dispersion at higher frequencies. For angular spread, the monotonic reduction with increasing frequency indicates more directional propagation at 15 GHz. Meanwhile, the systematic deviations of measured angular spreads from the 3GPP values demonstrate that the standard model fails to fully capture the specific propagation characteristics of the 8 and 15 GHz bands. In terms of MIMO capacity, the 15 GHz channel slightly outperforms 8 GHz in both LOS and NLOS scenarios due to more concentrated multipath energy and larger dominant singular values. Overall, higher frequencies exhibit greater directionality and environmental sensitivity, while lower frequencies offer broader multipath distributions, providing insights for the design and modeling of future multi-band 6G MIMO systems.


\section*{Acknowledgment}

This work was supported by National Natural Science Foundation of China (62341128, 62201086, 62525101), National Key Research and Development Program of China (2023YFB2904805), Beijing Municipal Natural Fund (L243002) in part and Beijing University of Posts and Telecommunications-China Mobile Research Institute Joint Innovation Center.

% \IEEEtriggeratref{7}
% \IEEEtriggercmd{\enlargethispage{-20cm}}


%\bibliographystyle{IEEEtran}

\begin{thebibliography}{00}
\bibitem{wrc23}ITU, “Final Acts of the World Radiocommunication Conference (WRC-23),” Dubai, United Arab Emirates, Nov.–Dec. 2023.
\bibitem{FR3} M. Ying et al., “Upper Mid-Band Channel Measurements and Characterization at 6.75 GHz FR1(C) and 16.95 GHz FR3 in an Indoor Factory Scenario,” ICC 2025 - IEEE International Conference on Communications, Montreal, QC, Canada, 2025, pp. 3303-3308.
\bibitem{1}Zhang J, et al., “Channel measurement, modeling, and simulation for 6G: A survey and tutorial,” arXiv preprint arXiv:2305.16616, 2023.
\bibitem{2}J. Zhang, et al., “New Mid-Band for 6G: Several Considerations from the Channel Propagation Characteristics Perspective,” in IEEE Communications Magazine, pp. 1-6, 2024.
\bibitem{6G} W. Saad, M. Bennis and M. Chen, “A Vision of 6G Wireless Systems: Applications, Trends, Technologies, and Open Research Problems,” in IEEE Network, vol. 34, no. 3, pp. 134-142, May/June 2020.
\bibitem{haiyang} H. Miao et al., “Measurement-Based Massive MIMO Channel Characterization in 6 GHz Band for 6G,” 2024 IEEE Wireless Communications and Networking Conference (WCNC), Dubai, United Arab Emirates, 2024, pp. 1-6.
\bibitem{haiyang2} H. Miao et al., “Sub-6 GHz to mmWave for 5G-Advanced and Beyond: Channel Measurements, Characteristics and Impact on System Performance,” in IEEE Journal on Selected Areas in Communications, vol. 41, no. 6, pp. 1945-1960, June 2023.
\bibitem{haiyang3} Z. Ying, T. Jiang, P. Tang, J. Zhang and L. Tian, “Analysis Of Delay Characteristics At 4.9 Ghz And 28 Ghz In An Indoor Industrial Scenario,” 2020 14th European Conference on Antennas and Propagation (EuCAP), Copenhagen, Denmark, 2020, pp. 1-5.
\bibitem{peter} M. Peter, R. Weiler, B. Göktepe, W. Keusgen, and K. Sakaguchi, “Channel Measurement and Modeling for 5G Urban Microcellular Scenarios,” Sensors, vol. 16, no. 8, p. 1330, Aug. 2016.
\bibitem{wang}R. Wang, “Enabling Super-Resolution Parameter Estimation for Mm-wave Channel Sounding,” IEEE Trans. Wireless Commun., vol. 19, no. 5, pp. 3077–3090, May 2020.
\bibitem{sage}J. A. Fessler and A. O. Hero, “Space-alternating generalized expectation-maximization algorithm,” in IEEE Transactions on Signal Processing, vol. 42, no. 10, pp. 2664-2677, Oct. 1994.
\bibitem{tangpan}P. Tang, “Channel Measurement and Path Loss Modeling From 220 GHz to 330 GHz for 6G Wireless Communications,” China Commun., vol. 18, no. 5, pp. 19–32, May 2021.
\bibitem{zhang}G. Zhang, “Millimeter-Wave Channel Characterization in Large Hall Scenario at the 10 and 28 GHz Bands,” in Proc. 13th Eur. Conf. Antennas Propag. (EuCAP), Mar. 2019, pp. 1–4.
\bibitem{3gpp} 3GPP, “3GPP TR 38.901 V14.1.1 Technical Report: Study on Channel Model for Frequency Spectrum Above 6 GHz (Release 14.1.1),” 2017.


\bibitem{capacity} C. E. Shannon, “A mathematical theory of communication,” in The Bell System Technical Journal, vol. 27, no. 3, pp. 379-423, July 1948.
\end{thebibliography}

\end{document}